\documentclass[reprint,
twocolumn,
 amsmath,amssymb,
 aps,pre
]{revtex4-2}

\usepackage{graphicx}
\usepackage{dcolumn}
\usepackage{bm}

\usepackage[dvipsnames]{xcolor}
\newcommand{\CHANGE}[1]{{\color{black}{{#1}}}}

\begin{document}

\preprint{arXiv:2502.05166}

\title{Stirring supercooled colloidal liquids at the particle scale}

\author{Piotr Habdas}
\email{phabdas@sju.edu}
\affiliation{Department of Physics, Saint Joseph's University, Philadelphia, PA 19131}

\author{Eric R. Weeks}%
\email{erweeks@emory.edu}
\affiliation{Department of Physics and Astronomy, Emory University, Atlanta, GA 30322}

\date{\today}

\begin{abstract}
We study the decay of tangential velocity profiles with distance from a local disturbance in hard-sphere colloidal suspensions as the colloidal glass transition is approached. The disturbance, generated by a dimer of superparamagnetic particles rotated by an external magnetic field, enables a precise characterization of the system's response through confocal microscopy and tracking of individual particle dynamics. The tangential velocity profiles exhibit nearly exponential decay with distance. As particle density increases toward the colloidal glass transition, the characteristic length scale derived from exponential fits grows.  We also observe that the colloidal particles slip against the rotating dimer, with less slip in samples which are closer to the glass transition.

\end{abstract}

\maketitle


\section{\label{sec:intro}Introduction}

When some liquids are rapidly cooled, they can become a glass.  In these circumstances, the viscosity of the liquid increases by many orders of magnitude, with little or no apparent change in the structure \cite{angell95,ediger96,angell00q,debenedetti01,dyre06,lubchenko07,chandler10,berthier11rmp}. Understanding the causes of the glass transition is a long-standing puzzle of condensed matter physics.

One approach to addressing questions about the glass transition is to use colloids as a model system.  Colloidal suspensions are composed of micron-sized particles in a liquid, and the particles diffuse due to Brownian motion.  The diffusion constant decreases rapidly as the volume fraction $\phi$ is increased toward $\phi_g \approx 0.58$, which has been identified as the colloidal glass transition point \cite{pusey86,hunter12rpp}.  Colloids have a long history of being used to study questions related to the glass transition \cite{pusey86,vanmegen94,segre95,vanmegen98,kasper98,schweizer03}.  \CHANGE{Simulations have shown that Brownian dynamics (true for colloidal experiments) lead to similar results as ballistic dynamics (realistic for molecular glasses) \cite{gleim98,tokuyama03,szamel04,lopezflores12,tokuyama07,berthier07}.  Moreover, hydrodynamic interactions are not believed to be important for time scales related to rearrangements \cite{weeks07cor,hunter12rpp,thorneywork15}.  The value of $\phi_g$ measured in colloidal samples matches that of simulations of hard spheres without hydrodynamics \cite{pusey09,ikeda12,ikeda13}.}
Due to the particle size and diffusion time scales, colloids are well suited for optical microscopy studies of the glass transition \cite{marcus99,habdas02,kegel00,weeks00}.

One of the first experiments studying a colloidal glass via optical microscopy was by van Blaaderen and Wiltzius in 1995 \cite{vanblaaderen95}.  They used a confocal microscope to observe the positions of $10^4$ colloidal particles in three dimensions, and found that there was no structural correlation length scale.  The structure was consistent with computer simulations of random close packed structures, and also agreed with simulations of glasses which also did not find any structural correlation length scales \cite{ernst91}.  This lack of obvious structural changes is one of the key features that distinguishes the glass transition from regular phase transitions.  This then is a puzzle: why does the viscosity of a sample grow dramatically near the glass transition (whether molecular liquids \cite{hecksher08} or colloids \cite{segre95,cheng02,russel13}) if the structure is unchanged?  One approach has been to look for other length scales that might change, such as dynamical heterogeneity (groups of particles that diffuse together) \cite{marcus99,kegel00,weeks00,weeks07cor,singh23,yun24}, identifying subtle particle packing structures \cite{cianci06ssc,royall15,hallett18}, looking at the effects of confining samples to narrow spaces \cite{nugent07prl,sarangapani08,sarangapani11,sarangapani12,zhang16,villada-balbuena22}, pinning subsets of particles in colloidal samples \cite{gokhale14,himanagamanasa15}, and employing local perturbations to disturb samples \cite{habdas04,anderson13,li20,khair06,gazuz09,zia14,puertas14,cicuta07sm,wilson09,li22}.  Some of these experiments complement theoretical ideas that postulate the existence of subtle well-packed random structures in glassy samples \cite{kivelson94,tarjus95,lubchenko07,kawasaki07}.


A different question has been studied in jammed materials.  Jammed materials are amorphous solids of any sort, perhaps including glasses but also including foams, emulsions, colloids, and granular materials \cite{liu98,cipelletti02,liu10,vanhecke10,behringer18,delgado23}.  These materials are solid-like when their constituent particles (droplets, bubbles, grains) are packed together at a high volume fraction, analogous to  colloidal glasses. 
Many experiments have looked at how these materials respond when being sheared \cite{howell99,mueth00,losert00,debregeas01,Lauridsen02,Wang06,coussot02mri,huang05,rodts05,lauridsen04}.  In many cases, the velocity profile decays exponentially away from a moving wall, for example, in granular systems \cite{howell99,mueth00,losert00} and bubbles confined between two plates \cite{debregeas01,Lauridsen02,Wang06}.  In some specific cases, the velocity profile discontinuously jumped to zero at a transition between a flowing region and a jammed region; this was observed in emulsions and colloids \cite{coussot02mri}, wet granular systems \cite{huang05}, three-dimensional foams \cite{rodts05}, bubble rafts \cite{lauridsen04,Wang06}, and lipid monolayers \cite{choi11}.

In this paper, we employ a microscopic stir bar to shear dense colloidal suspensions on the particle scale.  Specifically, we rotate dimers composed of a pair of superparamagnetic particles immersed in nearly hard-sphere colloidal suspensions.  The samples cover a range of volume fractions, with the highest volume fraction ($\phi \approx 0.56$) fairly close to the colloidal glass transition ($\phi_g = 0.58$).  We observe exponentially decaying velocity profiles, with the decay length growing by a factor of three as the glass transition is approached.  This provides a new observation of a growing length scale near the glass transition, and complements prior studies of sheared jammed materials.  Our experiments show that the colloidal sample engages more strongly with the rotating dimer as the glass transition is approached, suggesting that the sample is indeed  composed of tightly interlocked subtle structure.

\section{\label{sec:expt}Experimental Methods}

Our colloidal suspensions are made of poly-(methylmethacrylate) particles, which are sterically stabilized by a thin layer of poly-12-hydroxystearic acid \cite{antl86}, and are the same as used in our prior work \cite{habdas04,anderson13}.  The particles have a radius $a=1.10$ $\mu$m, a polydispersity of $\sim$5\%, and are dyed with rhodamine.  

The colloidal particles are suspended in a mixture of cyclohexylbromide/{\it cis}- and {\it trans}- decalin which nearly matches both the density and the index of refraction of the colloidal particles.  The control parameter is the volume fraction $\phi$, the amount of volume occupied by the solid particles.  In this mixture, the particles have hard sphere cores and also a slight charge \cite{hernandez09}; from prior work with similar samples in our laboratory, we know that these samples can have coexistence between a liquid at $\phi_{\rm freeze}=0.43$ and a crystal at $\phi_{\rm melt}=0.49$, and a glass transition at $\phi_{\rm glass} \approx 0.58$ \cite{hernandez09,habdas04}.  We add a small quantity of superparamagnetic beads (Thermo Fisher Scientific) with a radius of 2.25 $\mu$m.  We do not observe sticking of the colloidal particles to the magnetic beads.

\begin{figure}[!ht]
\includegraphics[scale=0.5]{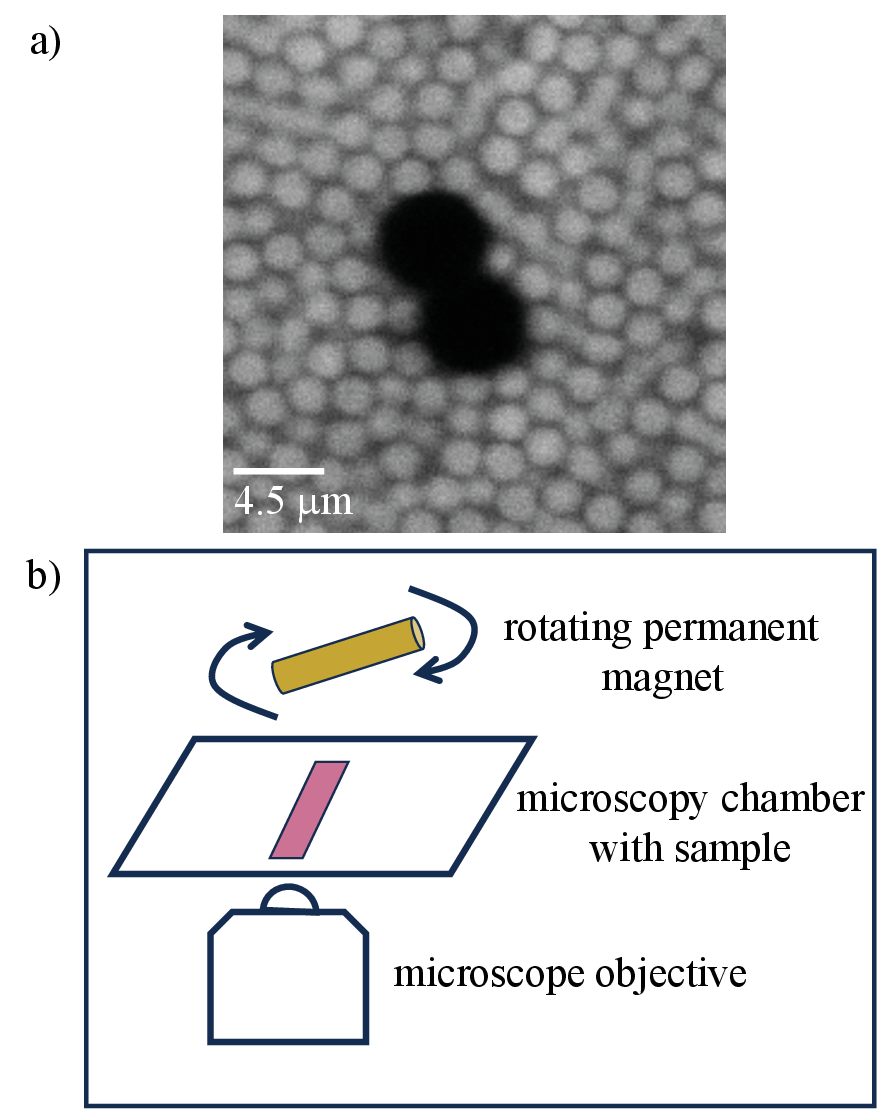}
\caption{(a) A snapshot of a dimer in a colloidal suspension, $\phi=0.558$.  (b) A schematic of the experimental setup.}
\label{dimer} 
\end{figure}

The samples we study are all at high volume fractions, $\phi \geq 0.50$.  Accordingly, the samples can form colloidal crystals after some period of time \cite{pusey86,hernandez09}.  Before taking data, we stir the sample with an air bubble to disrupt any crystalline regions that may be present, and then wait at least 20 minutes before taking data.  After stirring, we look for pairs of magnetic beads, i.e., dimers.  We only take data on dimers that are sufficiently far from chamber walls (at least 35 $\mu$m) and other magnetic beads (at least 150 $\mu$m).  One such dimer is shown in Fig.~\ref{dimer}(a).  If the sample does not have significant crystallization after taking a data set, we frequently continue collecting data using the same dimer without any sample stirring.  \CHANGE{Fortunately, crystallization is reasonably slow, especially at lower volume fractions \cite{harland97,auer01,gong07}; it typically takes 1-3 hours before we observe crystalline regions within our field of view.}  We also repeat experiments with the same sample but with different dimers.

A permanent Neodymium magnet is fixed to a stepper motor just above the objective of the microscope and the sample, as shown in Fig.~\ref{dimer}(b).  The motor allows us to rotate the permanent magnet at various rotational speeds ranging from 1.25 rev/h to 12.5 rev/h; we primarily study two rotation speeds, 5.0 and 12.5 rev/h.  The presence of the magnet above the slide also exerts a slight upward force on the dimer.  We position the magnet at a height where the magnetic force balances the downward gravitational force on the dimer, so that the dimer stays in focus during the duration of our experiment.  We let the dimer rotate for at least one full rotation prior to collecting data. 

With a confocal microscope, we rapidly acquire images of area $80 \times 80$~$\mu$m$^2$, containing several hundred particles.  We use a $100\times$ objective with numerical aperture 1.4.  The dimers are not fluorescent and appear black on the background of the dyed colloidal particles [Fig. \ref{dimer}(a)].  Using traditional tracking techniques we follow the position of each colloidal particle as well as the dimer in time \cite{Crocker96}.  The uncertainty of particle positions is approximately $\pm 50$~nm for our images.

We also take three-dimensional (3D) data sets on the same samples, typically scanning a volume $80 \times 80 \times 20$~$\mu$m$^3$, and analyzing the data using standard techniques to identify particle positions in 3D \cite{Crocker96,dinsmore01}.  The volume fraction $\phi$ is calculated by summing the volume of all colloidal particles whose centers lie within the observational volume, and dividing by the volume of the observation region.  The uncertainty of the mean particle radius ($\pm 0.02$~$\mu$m) leads to a systematic uncertainty of our volume fraction of 3\%; that is, a nominal volume fraction of $\phi=0.500$ lies within the range $0.485 - 0.515$ \cite{poon12}.  However, the relative uncertainty of our volume fractions is much smaller (less than 0.5\% uncertainty) so we report our volume fractions to three significant digits to allow comparison of relative volume fractions. 

To quantify the strength of the deformation of the system as the glass transition is approached we calculate the modified Peclet number $Pe^*$.  The Peclet number is a ratio of two time scales, $\tau_{\rm diff}$ characterizing the unforced diffusive motion of the colloidal particles, and $\tau_{\rm rot}$ characterizing the magnetic bead rotation.  The Brownian time $\tau_{\rm diff}=a^2/2D_\infty$ is the time it takes for colloidal particles to diffuse a distance equal to their radius $a$ and is based on their asymptotic diffusion constant $D_\infty$ which depends on the volume fraction $\phi$.  It is this use of the $\phi$-dependent diffusivity that is the reason we term this the ``modified'' Peclet number (as opposed to using $D_0$, the diffusion constant for a dilute sample).  $D_\infty$ decreases as the colloidal glass transition is approached due to increased particle crowding \cite{vanmegen94,vanmegen98,weeks02}.  The magnetic dimer time scale $\tau_{\rm rot} = 1/(2 \pi \Omega)$ is the time scale for the magnetic dimer to make one full rotation.  The modified Peclet number is then $Pe^* = \tau_{\rm dif} / \tau_{\rm rot}$ and ranges from approximately 5 - 450 for these experiments.  Thus, in these experiments, the forced rotation of the magnetic dimer is much more significant than the Brownian motion of the surrounding colloidal particles.  As the magnetic dimer rotates, it pushes colloidal particles out of the way, plastically deforming the sample.  The particle diffusive motion is too slow for the sample to equilibrate to the dimer orientation.
  
\section{\label{sec:results}Results and Discussion}

\subsection{Mean velocity profiles}

\begin{figure}[!tbh]
\includegraphics[width=0.95\columnwidth]{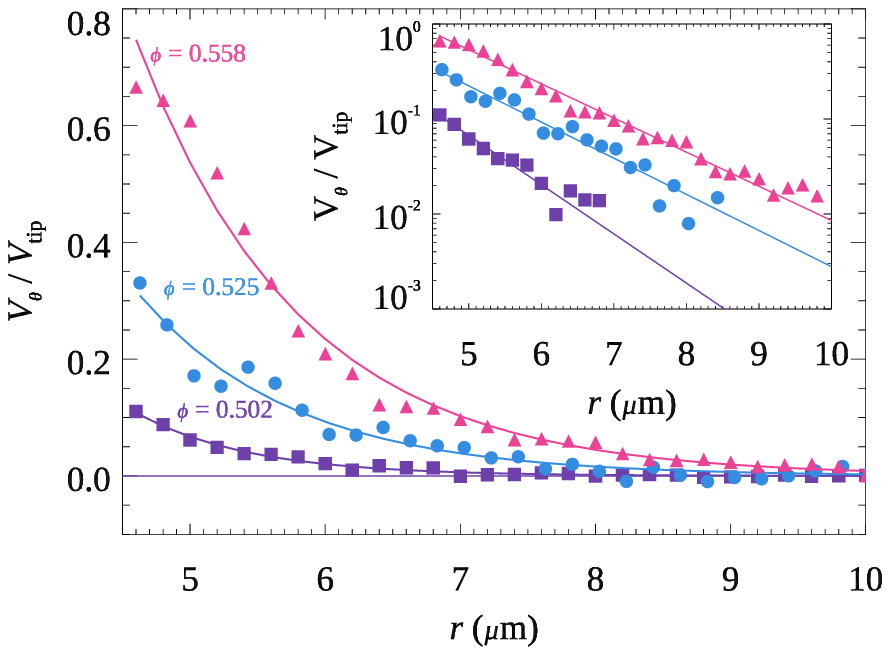}
\caption{Azimuthal particle velocity (normalized by the $V_{\rm tip}$ of the dimer) as a function of distance from the dimer center for volume fractions as labeled on the plot. 
 Solid lines are fits to $A e^{-(r-r_{\rm tip})/\lambda}$.
 $Pe^*$ $\approx$ 250 with $A$ = 0.122 $\pm$ 0.005 and $\lambda$ = 0.84 $\pm$ 0.5 $\mu$m (squares), $Pe^*$ $\approx$ 315 with $A$ = 0.35 $\pm$ 0.02 and $\lambda$ = 1.14 $\pm$ 0.07 $\mu$m (circles), and $Pe^*$ $\approx$ 412 with $A$ = 0.81 $\pm$ 0.02 and $\lambda$ = 1.21 $\pm$ 0.05 $\mu$m (triangles). For all three data sets, $\Omega$ = 12.5 rev/h and $V_{\rm tip}$ = 0.098 $\mu$m/s.  The inset shows the same data on a semi-log plot.}
\label{Vtheta} 
\end{figure}

 We calculate the average azimuthal velocity of the colloidal particles $V_\theta$ as an azimuthal distance traveled by the colloidal particles in the time that takes the dimer tip to rotate a colloidal particle radius; that is, using a time scale $\Delta t = a/(\Omega r_{\rm tip})$.  The average is calculated over the entire experimental duration of about 45 minutes, over which the dimer rotates 3.8 - 9.4 revolutions depending on $\Omega$. 

The average azimuthal velocity $V_\theta$, scaled by the tangential velocity of the dimer tip $V_{\rm tip}$, is plotted as a function of distance from the dimer center $r$ in Fig.~\ref{Vtheta}, for selected volume fractions and dimer rotational angular speed of $\Omega$ = 12.5 rev/h (0.0218 rad/s).  The plot starts at $r = r_{\rm tip} = 4.5$~$\mu$m.  We see that the maximum $V_\theta$/$V_{\rm tip}$ at $r_{\rm tip}$ is less than one, revealing that there is a slip between the dimer and the colloidal particles.  For larger $\phi$, $V_\theta$/$V_{\rm tip}$ at the dimer tip increases, indicating that with increased particle crowding there is less slip between the dimer tip and the colloidal particles.

The inset of Fig.~\ref{Vtheta} shows that the azimuthal velocity decays exponentially as a function of the distance from the dimer. 
Far from the dimer, the particles primarily move via Brownian motion within the cages formed by their neighbors \cite{weeks02}.  This motion is random and so should average to zero; but with a finite amount of data, the average will not quite be zero, but will fluctuate around zero.  This residual noise level is about $10^{-3} - 10^{-2}$, with the value depending on how long a given movie is, particle tracking details, and the volume fraction.  Closer to the dimer, the data are above this noise level.  We note that the velocity profiles are independent of the choice of $\Delta t$. To fit the exponential decay, we estimate the critical distance $r_{\rm tip} + r_c$ where $V_\theta$/$V_{\rm tip}$ reaches the noise level, i.e. $r$ at which $V_\theta$/$V_{\rm tip}$ reaches average $V_\theta$/$V_{\rm tip}$ far from the dimer ($r >$ 20 $\mu$m) plus two standard deviations.  Fitting just the data for $r_{\rm tip} \leq r \leq r_{\rm tip}+r_c$, we find that the decay of the normalized azimuthal velocity is well fitted by an exponential function $A e^{-(r-r_{\rm tip})/\lambda}$ (solid lines in Fig. \ref{Vtheta}).  As a comparison, for a sphere rotating in a normal Newtonian fluid, the flow field decays as $1/r^2$ \cite{currie74}.  We note that a simple power law function does not fit our data well, even though some similar experiments observed a power law behavior in quasi-two-dimensional foams \cite{Lauridsen02,Wang06} and in several soft amorphous solids \cite{coussot02mri}; the latter study focused on the macroscopic (cm-scale) velocity profile, orders of magnitude above the constituent particle size.

\begin{figure}[!tbh]
\includegraphics[width=0.95\columnwidth]{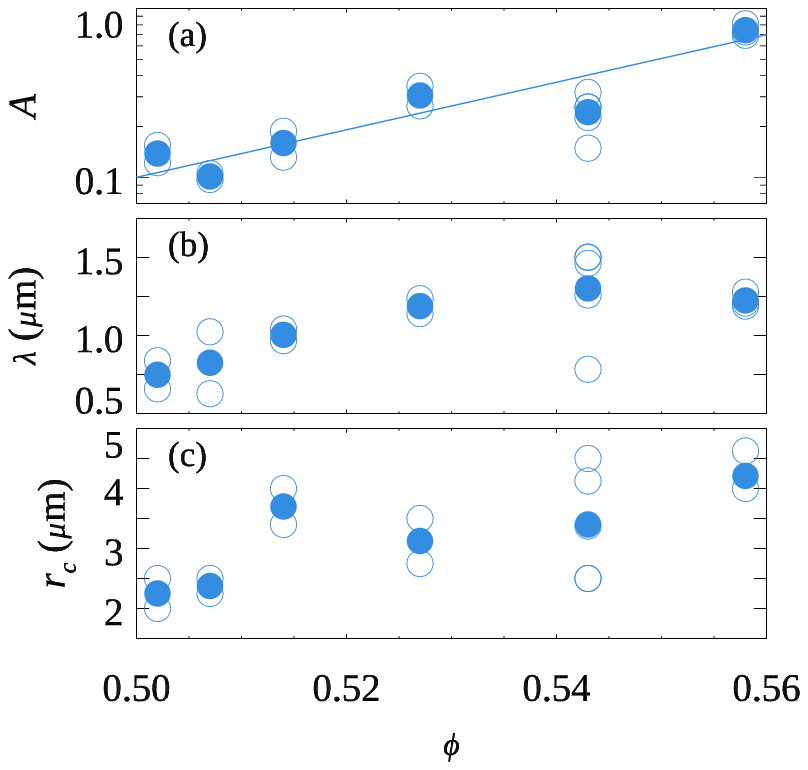}
\caption{\label{rclambda} (a) Amplitude $A$ (open circles) and average $A$ (closed circles).  The solid line is a guide to the eye. (b) Decay length $\lambda$ (open circles) and average $\lambda$ (closed circles), both from the exponential fits of $V_\theta (r)/V_{\rm tip}$, as a function of volume fraction $\phi$ for $\Omega$ = 12.5 rev/h. (c) Critical length scale $r_c$ (open circles) and average $r_c$  (closed circles) obtained from the $V_{\theta}(r)/V_{\rm tip}$ profiles as a function of volume fraction $\phi$ for $\Omega$ = 12.5 rev/h, Pe$^* \sim 100 - 250$.}
\end{figure}

As is apparent from the inset to Fig.~\ref{Vtheta}, at larger $\phi$ the decay length $\lambda$ is larger.  The fitting data from all experiments with $\Omega = 12.5$ rev/h are shown in Fig.~\ref{rclambda}, confirming this trend, albeit with a fair bit of variability within each sample.  The variability of $\lambda$ may be due to the intrinsic spatial heterogeneity of the glassy samples.  The values of $A$ and $\lambda$ are not sensitive to the choice of $r_c$ (which determines the fitting range).

Not surprisingly, the amplitude $A$ of the velocity profiles increases with volume fraction [Fig.~\ref{rclambda}(a)].  Increasing $A$ indicates that the dimer has less slip with the sample at higher $\phi$.  Similarly, the average decay length grows from about 0.7~$\mu$m to 1.3~$\mu$m, so is the same scale as the particle radius $a = 1.10$~$\mu$m [Fig.~\ref{rclambda}(b)].  This is significantly shorter than other length scales which have been seen in colloidal glasses, such as the length scales associated with cage rearrangements ($3a -7a$, Ref.~\cite{weeks07cor}) and length scales associated with confinement effects ($7a-30a$, Refs.~\cite{sarangapani11,karmakar09,yun24,villada-balbuena22,nugent07prl}).  Another length scale of interest is the critical distance $r_c$, where $V_\theta$/$V_{\rm tip}$ reaches the noise level.  This is plotted in Fig.~\ref{rclambda}(c), and likewise grows with increasing $\phi$.  This is essentially a consequence of $A$ and $\lambda$, which determine when $V_\theta$/$V_{\rm tip}$ reaches the noise level.

\begin{figure}[!tbh]
\includegraphics[width=0.95\columnwidth]{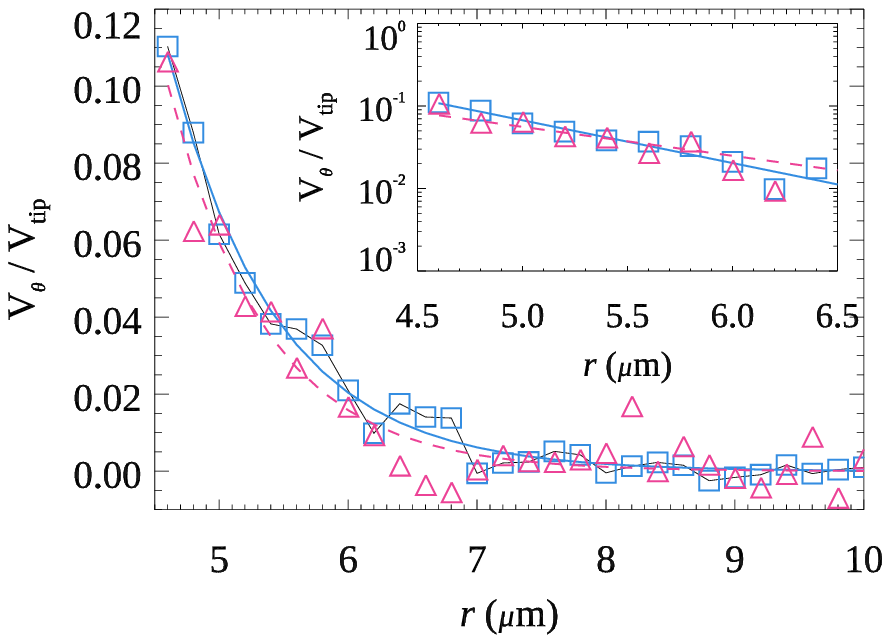}
\caption{\label{VthetaSpeeds} $V_\theta$/$V_{\rm tip}$ as a function of distance from the dimer center $r$ for $\phi$ = 0.502 and $\Omega$ = 5 rev/h, $Pe^*$ $\approx$ 100 (triangles) and $\Omega$ = 12.5 rev/h, $Pe^*$ $\approx$ 250 (squares).  Lines are fits to $A e^{-(r-r_{\rm tip})/\lambda}$.  For $\Omega$ = 5 rev/h (dashed line): $A$ = 0.122 and $\lambda$ = 0.838 $\mu$m.  For $\Omega$ = 12.5 rev/h (solid line): $A$ = 0.115 and $\lambda$ = 0.757 $\mu$m.  Note that $r = 0$ $\mu$m corresponds to the center of the dimer; the plot starts at $r=4.5$~$\mu$m which corresponds to the tips of the dimer.  The inset shows the same data on a semi-log plot.}
\end{figure}

We also examine the dependence of the $V_\theta$/$V_{\rm tip}$ decay on the rotating speed of the dimer.  Figure~\ref{VthetaSpeeds} shows $V_\theta$/$V_{\rm tip}$ for $\phi$ = 0.502 and two different rotating speeds of the dimer $\Omega$ = 5 rev/h (triangles) and $\Omega$ = 12.5 rev/h (squares).  Interestingly, the decay of the normalized azimuthal particle velocity is independent of rotational speed of the dimer; for both data sets, they reach the noise level around $r_{\rm tip}+r_c$ $\approx 6.5$~$\mu$m.  This suggests that for this Peclet number range (Pe$^* \sim 100 - 250$), the behavior is in the high Peclet number limit and independent of Pe$^*$.  This is consistent with a prior experiment which found an independence of the response length scale to the probing strength, where the colloidal sample responded to a local radial expansion force \cite{li20}.  Figure~\ref{VthetaSpeeds} also shows that it is harder to measure $\lambda$ for the slower rotation rate, where the particle velocities are slower and the data are more noisy (triangle symbols in the figure).


\subsection{Discontinuous velocity transition?}

As described above, our data for $V_\theta$ are well fit by exponential decay.  This implies that at any distance $r$, there is some amount of motion induced by the rotating dimer, even if well below the noise.  An alternative hypothesis is that there could be a discontinuous transition between a sheared region and a stagnant region, as has been seen before in other soft matter experiments \cite{coussot02mri,Lauridsen02,Wang06,choi11}.  The concept is that some soft materials have a yield stress: they do not flow unless the applied stress exceeds a minimum value \cite{joshi18,cerbino23,divoux24}.  In our situation, the logic is that we rotate the dimer at a high Peclet number (Pe$^* \geq 5$):  the dimer rotates faster than the sample can equilibrate, and in this situation, the sample possesses a yield stress \cite{habdas04}.  Small stresses below the yield stress cause the sample to strain elastically, but due to the caging of the colloidal particles, they do not rearrange if the stress is small \cite{anderson13}.  That is, the dimer rotates at a fixed rate, exerting a controlled strain on the sample.  The stress is whatever is necessary for this rotation, and near the dimer, the applied stress exceeds the yield stress, causing the sample to flow.  This is the motion shown in Fig.~\ref{Vtheta}.  However, the torque caused by the dimer is distributed over a larger area at larger $r$, causing the exerted stress to decrease until a distance at which it falls under the yield stress.  Beyond that distance, the sample will deform elastically, and the elastic stress in that far-field region will balance the viscous stress exerted by the flowing region.  This implies that at some distance, $V_\theta$ will drop discontinously to zero; as noted above, similar stagnant zones have been observed in previous experiments on different soft materials \cite{coussot02mri,Lauridsen02,Wang06,choi11}.

Given the noise in the data for $V_\theta / V_{\rm tip} \lesssim 10^{-2}$, our data cannot resolve this discontinuous transition.  Another possibility is that the sample continues to shear at all distances $r$, with velocities $V_\theta$ below our ability to resolve at large $r$.  This behavior for example was seen in an experiment which studied sheared granular material \cite{losert00}, for which the velocity profile decayed exponentially away from the moving surface.  In that experiment, there was always a chance that a faraway particle could be disturbed from its position and move, so it was plausible that there was not a discontinuous transition to a stagnant region.  Likewise, it is certainly plausible that in our conjectured stagnant zone, that occasionally a particle will be on the verge of a cage rearrangement \cite{weeks02} and the viscous stress could cause that rearrangement to occur.  These dimer-influenced motions would be biased in the direction of the rotating dimer and thus result in $V_\theta > 0$.  We speculate that if such displacements occur, they are on a much smaller scale than the exponential decay seen in Fig.~\ref{Vtheta}, and so at the very least, the character of the motion changes qualitatively at the transition from flowing to elastic.

To be clear, our data do not see a transition from $V_\theta(r)$ having exponential decay to any other behavior, whether a discontinuous drop to $V_\theta=0$ or merely a crossover to a different functional form.  To observe such possibilities would require taking a much larger quantity of data.  Given the character of Brownian motion, if we had four times as much data, we would be able to reduce the noise level in the far field by a factor of 2.  However, with an exponentially decaying $V_\theta(r)$, this extends the range of visibility of the velocity profile only a short distance.  Moreover, our experiments are limited in duration by a slow vertical drift of the dimer and by crystallization of the sample.  We note the possibility of a qualitative change of $V_\theta(r)$ because we believe that it is quite plausible even if we cannot observe this behavior.  Any location where a qualitative change occurs is at a distance farther away than $r_{\rm tip}+r_c$.

\subsection{Instantaneous response}

\begin{figure}[!tbh]
\includegraphics[scale=0.3]{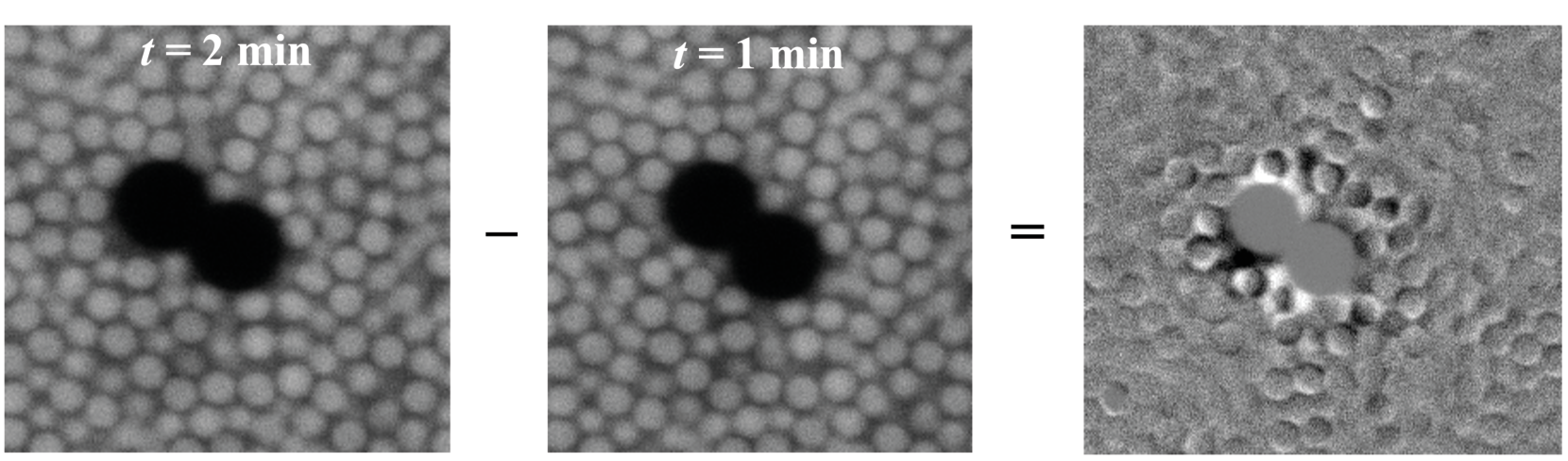}
\caption{\label{mydiff} Two snapshots of a rotating dimer taken at time interval of one minute $\phi$ = 0.527, $\Omega$ = 1.25 rev/h, rotation angle $\Delta \Theta = 7.5^\circ$ counterclockwise, and $Pe^*\approx 6$.  The third image is the subtraction of the second image from the first.  The black edge of the dimer indicates the direction the dimer is rotating, whereas for the colloidal particles, the white edges indicate their direction of motion.  Gray is where the two images are similar, indicating no motion.
}
\end{figure}

We have thus far focused on the time- and angle-averaged behavior.  However, the sample is composed of discrete particles and so it is natural that the behavior at any instant in time is spatially heterogeneous.  Evidence of this can be seen when we take two images at different times and subtract them from each other, as shown in Fig.~\ref{mydiff}.  Black and white indicate the difference and hence motion of colloidal particles, whereas gray indicates no motion.  Naturally, the most motion occurs in the vicinity of the rotating dimer.  However, the response is not rotationally symmetric; there are ``chains'' of particles that move together at this moment in time, and some particles are moving in the radial direction rather than purely in the $\theta$ direction.

\begin{figure}[!tbh]
\includegraphics[width=0.95\columnwidth]{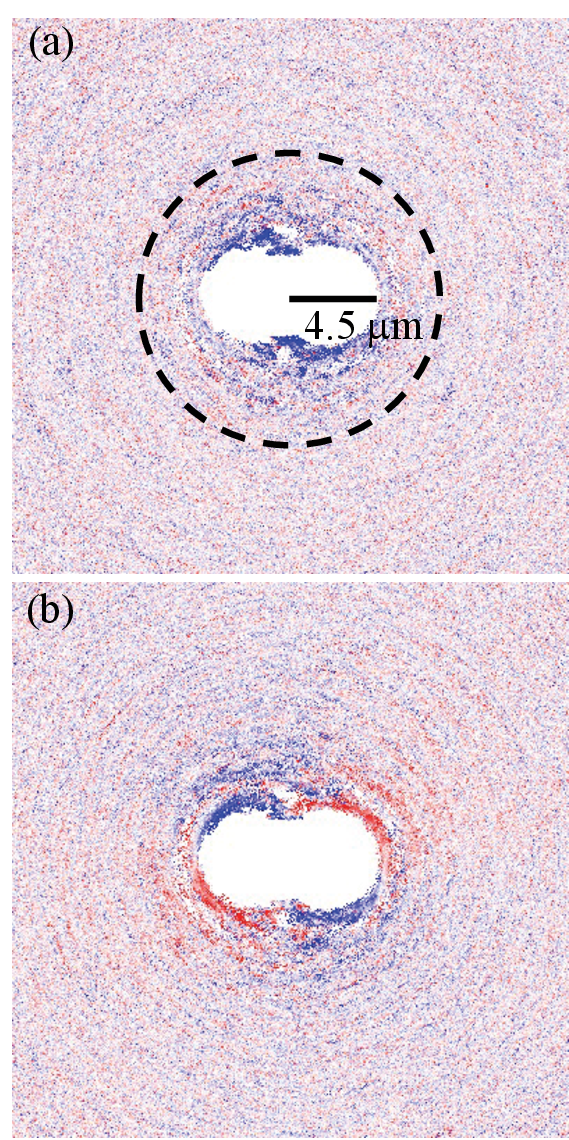}
\caption{\label{AverageAzimuthalDisp} a) Average flow field in the tangential direction.  Blue color indicates motion in the counter clockwise direction and red color indicates motion in the clockwise direction.  The dashed circle has a radius equal to $r_c$ = 7.5 $\mu$m.  b) Average flow field in the radial direction.  Blue color indicates motion towards the dimer center and red color indicates motion away from the dimer center.  Sample with $\phi$ = 0.527, dimer rotational speed $\Omega$ = 12.5 rev/h counter clockwise, $Pe^*$ $\approx$ 315.}
\end{figure}


An alternate look at the motion comes not from fluctuations but from doing a temporal average of the motion of the colloidal particles without the angular average.  In particular, we compute the time-averaged displacement field of the colloidal particles in the co-rotating dimer reference frame. Figure~\ref{AverageAzimuthalDisp} presents the average flow field for a sample with $\phi=0.527$ and rotational speed $\Omega$ = 12.5 rev/h, with (a) showing the $V_\theta$ motion and (b) showing the $V_r$ motion.  The azimuthal motion behaves as expected:  $V_\theta$ is largest near the dimer.  There is a slight $\theta$ dependence, where the locations ``behind'' the rotating dimer tips are the darkest blue, indicating the largest displacements.  The radial motion [Fig.~\ref{AverageAzimuthalDisp}(b)] shows interesting behavior, where there is a strong angular dependence of the radial motion.  Recall that the dimer rotates counterclockwise.  The two leading edges of the rotating dimer have red patches in Fig.~\ref{AverageAzimuthalDisp}(b), indicating outward radial motion.  In contrast, the trailing edges of the rotating dimer have blue patches, indicating an inward radial motion.  Given that there is no net flux, the $\theta$-average of these motions must be zero.  The radial motion also oscillates slightly as a function of $r$, most likely because the particles form slightly structured layers around the dimer, similar to what happens near walls \cite{nugent07prl,desmond09}.  Particles ``in between'' the layers have radial motions that differ from the particles inside the layers.

\begin{figure}[!tbh]
\includegraphics[width=1.0\columnwidth]{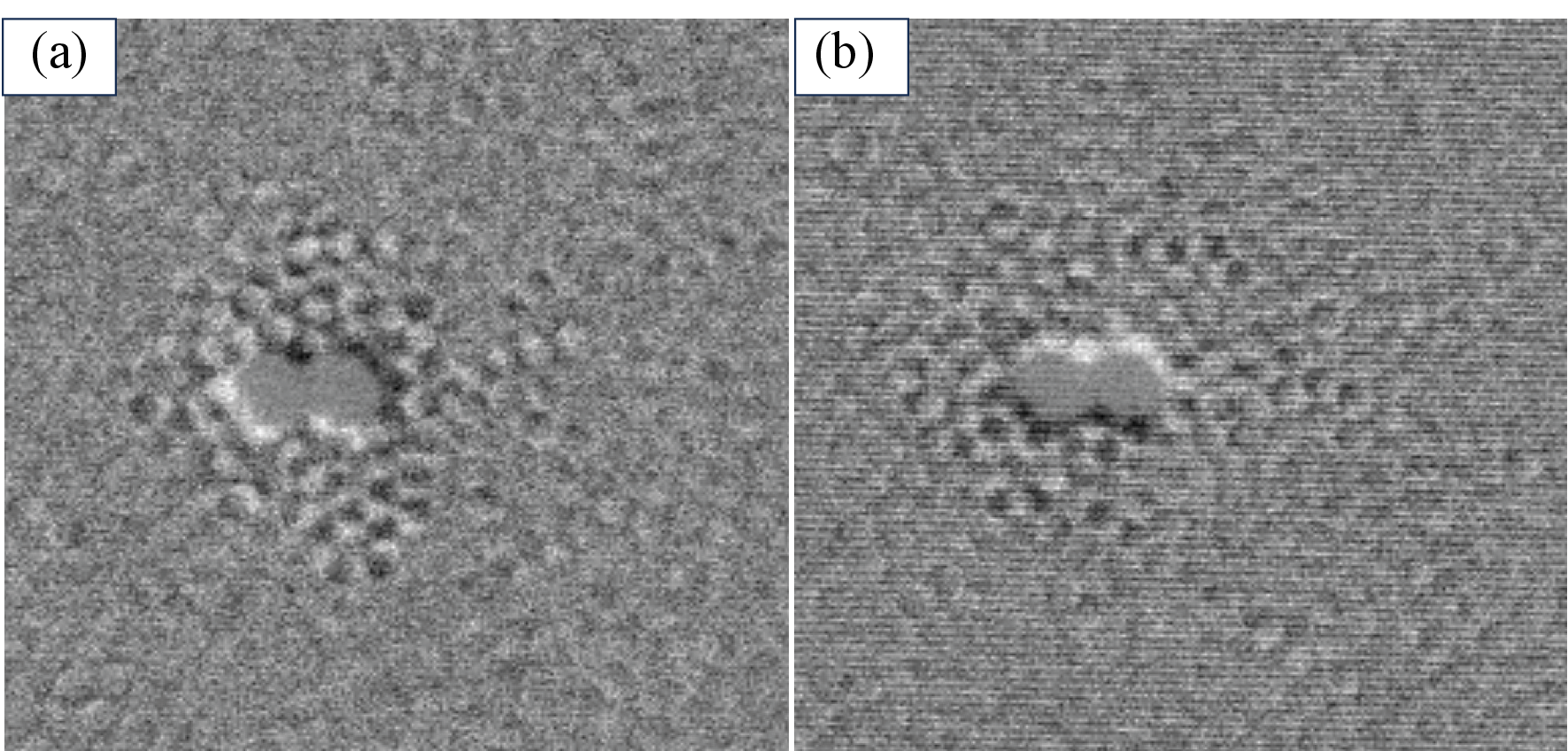}
\caption{\label{melting} \CHANGE{Two snapshots of a rotating dimer obtained by subtracting images with a time interval of $\sim145$ s which corresponds to the dimer rotating  $\Delta \Theta = 180^\circ$ clockwise. (a) $\Delta t = t_2 - t_1 \approx 906\ \mathrm{s} - 761\ \mathrm{s}$, (b) $\Delta t = t_2 - t_1 \approx 1053\ \mathrm{s} - 907\ \mathrm{s}$. $\phi = 0.558$, $\Omega = 12.5$ rev/h, $Pe^* \approx 412$.}}
\end{figure}

\CHANGE{One way to quantify the temporal heterogeneity of the motion is to compute not the mean motion (in other words, the velocity profiles shown in Fig.~\ref{Vtheta}) but rather the standard deviation of the motion (standard deviation of both azimuthal and radial velocity) as a function of $r$.  For all of our data sets, we find the standard deviation of the displacements to decay nearly exponentially to a nonzero plateau at large $r$ (data not shown).  The plateau height should be related to the mean square displacement for each sample at the given $\Delta t$.  Unfortunately, due to noise in the microscopy images, the plateau height varies from experiment to experiment; noisy images are known to increase the apparent level of the mean square displacement \cite{poon12}.  The decay to the plateau is relatively rapid and the standard deviation data are noisy, obscuring any trend with the volume fraction. Since the standard deviation in the radial direction is an order of magnitude greater than the standard deviation in the azimuthal direction, it indicates significant radial motion of the colloidal particles even though the {\it net} radial motion averages to zero.}

\CHANGE{The radial motion inward and outward evident in Fig.~\ref{AverageAzimuthalDisp}(b), and the discussion of the standard deviation of $v_{\theta}$ and $v_{r}$, raise the question of reversibility.  That is, it is possible that colloidal particles move periodically back and forth, returning to their original positions after a half-rotation of the dimer; but it is also possible that stirring locally ``melts'' the sample, allowing particles to move randomly in three dimensions around the dimer.  To clarify the behavior, Fig.~\ref{melting} presents two snapshots obtained in the same manner as Fig.~\ref{mydiff}, but with a time interval corresponding to a $180^\circ$ rotation of the dimer.  The black and white shading indicates particle motion. Both snapshots, taken at different times during the experiment, reveal that the colloidal particles move in random directions around the dimer, indicating that the system locally melts in the vicinity of the dimer.  Further evidence is that particles more frequently appear and disappear near the dimer, indicating motion in the direction perpendicular to the imaging plane.  Further inspection of the experimental movies, along with individual particle trajectories, confirms that the high standard deviation of the velocity components near the dimer corresponds to essentially random motion excited by the dimer, of course with the systematic mean $v_\theta(r)$ superimposed.}

\section{Conclusions}

We have examined the forced micro-stirring of dense colloidal suspensions at volume fractions close to the colloidal glass transition.  Our ``stir bars'' are dimers of two magnetic particles, with a dimer radius $r_{\rm tip} = 4.5$~$\mu$m approximately twice the diameter of the colloidal particles.  The stirring is rapid compared to the time scales of diffusion within these samples, with the modified Peclet number Pe$^*$ typically at least 100.  In the plastically deformed region near the dimer, the tangential velocity profiles exhibit exponential decay.  The decay length scale grows by approximately a factor of 3 as the volume fraction changes from $\phi = 0.50$ to 0.56, although in all cases this length scale is shorter than the diameter of the colloidal particles.  We also see that at the highest volume fraction, the dimer is maximally engaged with the colloidal sample: particles near the tip move with displacements quite similar to the tip motion.  In contrast, at lower volume fractions, there is slip at the tip, where the dimer tip moves substantially faster than the colloidal particles, by as much as a factor of 10.  This makes sense because there is more free volume for the colloidal particles to move slightly away from the dimer tip, allowing some of the shear to occur in an induced lower-volume fraction region.

Our observations add to prior work examining colloidal glasses for various length scales such as confinement \cite{sarangapani11,karmakar09,yun24,villada-balbuena22,nugent07prl} and dynamical heterogeneity \cite{marcus99,kegel00,weeks00,weeks07cor,singh23,yun24}.  Our observed decay lengths are significantly shorter than the prior observations.  In general, our results present a picture of colloidal samples that are hard to stir, in the sense that the disturbance decays quite quickly as a function of distance away from the stirring bar.  We conjecture that beyond a critical distance, the effect of the stirring may drop discontinuously to zero, with the sample only straining elastically beyond that distance.  The interesting twist is that for volume fractions closer to the glass transition, the region over which we can observe stirred motion grows due to the increased engagement of the colloidal particles with the rotating dimer.  This is consistent with theoretical ideas of the glass transition as interlocking regions of subtly structured well-packed particles which grow in size as the glass transition is approached \cite{kivelson94,tarjus95,lubchenko07,kawasaki07}.

\begin{acknowledgments}
We thank A. B. Schofield for synthesizing our colloidal particles.  We thank D. R. Nelson for the original inspiration for this project.  The initial data collection was supported by NASA (NAG3-2284) and subsequent data analysis was supported by the National Science Foundation (CBET-2002815, CBET-2333224).
\end{acknowledgments}

\CHANGE{
\section*{Data Availability}
}

\CHANGE{The data that support the findings of this article are openly
available \cite{dataonline}.}

\bibliography{eric}

\end{document}